\magnification=1200
\catcode`@=11 \def\b@lank{ }
\newif\if@simboli\newif\if@riferimenti\newif\if@incima\newif\if@bozze
\def\bozze{\@bozzetrue \immediate\write16{** printing eqnames}}
\newwrite\file@simboli\def\simboli{\immediate\write16{ Genera \jobname.SMB }
\@simbolitrue\immediate\openout\file@simboli=\jobname.smb
\immediate\write\file@simboli{Simboli di \jobname}}
\newwrite\file@ausiliario\def\riferimentifuturi{
\immediate\write16{ Genera \jobname.AUX }\@riferimentitrue\openin1 \jobname.aux
\ifeof1\relax\else\closein1\relax\input\jobname.aux\fi
\immediate\openout\file@ausiliario=\jobname.aux}
\newcount\eq@num\global\eq@num=0\newcount\sect@num\global\sect@num=0
\newcount\lemm@num\global\lemm@num=0
\newif\if@ndoppia\def\numerazionedoppia{\@ndoppiatrue\gdef\la@sezionecorrente{
\the\sect@num}}\def\se@indefinito#1{\expandafter\ifx\csname#1\endcsname\relax}
\def\spo@glia#1>{} \newif\if@primasezione
\@primasezionetrue \def\s@ection#1\par{\immediate\write16{#1}
\if@primasezione\global\@primasezionefalse\else\goodbreak
\vskip\spaziosoprasez\fi\noindent{\bf#1}\nobreak\vskip
\spaziosottosez\nobreak\noindent}
\def\eqpreset#1{\global\eq@num=#1\immediate\write16{ !!! eq-preset = #1 }     }
\def\eqlabel#1{\global\advance\eq@num by 1
\if@ndoppia\xdef\il@numero{(\la@sezionecorrente.\the\eq@num)}
\else\xdef\il@numero{(\the\eq@num)}\fi
\def\usa@getta{1}\se@indefinito{@eq@#1}\def\usa@getta{2}\fi
\expandafter\ifx\csname @eq@#1\endcsname\il@numero\def\usa@getta{2}\fi
\ifodd\usa@getta\immediate\write16
{ *** possibili riferimenti errati a \string\eqref{#1} !!!}\fi
\expandafter\xdef\csname @eq@#1\endcsname{\il@numero}
\if@ndoppia       \def\usa@getta{\expandafter\spo@glia\meaning
\la@sezionecorrente.\the\eq@num}\else\def\usa@getta{\the\eq@num}\fi
\if@simboli\immediate\write\file@simboli{  Equazione 
\usa@getta :  eqref.   #1}\fi
\if@riferimenti\immediate\write\file@ausiliario{
\string\expandafter\string\edef
\string\csname\b@lank @eq@#1\string\endcsname{\usa@getta}}\fi}
\def\eqref#1{\se@indefinito{@eq@#1}
\immediate\write16{ *****\string\eqref{#1} non definita !!!}
\if@riferimenti\relax\else\eqlabel{#1} @@?@@\fi \fi\csname @eq@#1\endcsname }
\def\autoeqno#1{\eqlabel{#1}\eqno\csname @eq@#1\endcsname\if@bozze
{\tt #1}\else\relax\fi}\def\autoleqno#1{\eqlabel{#1}
\leqno(\csname @eq@#1\endcsname)}\def\titoli#1{
\@incimatrue\nopagenumbers\xdef\prima@riga{#1}
\if@incima\voffset=+30pt\headline={\if\pageno=1{\hfil}\else\hfil{\sl 
\prima@riga}\hfil\folio\fi}\fi}\catcode`@=12

\def\frac#1#2{{#1\over {#2}}}
\def\diff#1#2{{\partial #1\over\partial #2}}
\def\di#1#2{{ d#1\over\ d #2}}
\def\pretitolo{\par\vskip .7cm\noindent} 
\def\postitolo{\par\nobreak\vskip .3cm\nobreak} 
\def\pre{\par\vskip .5cm\noindent} \def\post{\par\nobreak\vskip .2cm\nobreak} 
\def\capo{\par\noindent } \def\tre{{1\over 3}}\def\due{{1\over 2}}

\def\id#1#2{\delta ^{#1}_{#2}}

\def\Epsilon{{\cal E}}
\def\H{{\sqrt{H}}}
\def\hh{{\cal H}}
\def\const{\rm const.}
\def\X{\xi}
\def\a{R^0}
\def\al{\alpha}
\def\ala{\gamma}
\def\lu{\lim_{R,m\to 0}}
\def\s{{\cal S}}

\def\circa{\simeq}

\nopagenumbers
\
\vskip 1 truecm
\capo

\centerline{{\bf GRAVITATIONAL COLLAPSE}}\capo
\centerline{{\bf WITH NON--VANISHING TANGENTIAL STRESSES II:}}\capo
\centerline{{\bf A  LABORATORY FOR COSMIC CENSORSHIP EXPERIMENTS}}
\footnote{\ }{
Class. Quantum Gravity {\bf 15}, (1998), p. 3215--3228}

\vskip 3 truecm 
\capo
\centerline{{\bf Giulio Magli}}\capo
\vskip 1 truecm
\centerline{
Dipartimento di Matematica del Politecnico di Milano}\capo
\centerline{P.le Leonardo da Vinci 32, 20133 Milano, 
Italy}\capo
\centerline{e-mail magli@mate.polimi.it}
\vskip .2 truecm

\vskip 3 truecm\capo
{\sl 
The general
exact solution describing 
the dynamics 
of ani\-so\-tro\-pic elastic spheres 
supported only by tangential stresses
is reduced to a quadrature using 
Ori's mass--area coordinates.
This leads to 
the explicit construction of the
root equation governing the nature of the central singularity.
Using this equation,
we formulate and motivate on physical grounds
a conjecture on the nature of this singularity.
The conjecture 
covers a large sector of the space of initial data;
roughly speaking, it asserts that 
addition of a tangential stress 
cannot undress a covered dust singularity.
The root equation also allows us to analyze
the case of self--similar spacetimes and
to get some insight on the role of stresses 
in deciding the nature of the singularities 
in this case.
}

\vskip 3 truecm\capo

\vfill\eject 

\pageno=1

\titoli{G. Magli, Gravitational collapse with non--vanishing
tangential
stresses II.}

\pretitolo
{\bf I. Introduction.}
\postitolo

The present status of research in black hole formation
and cosmic censorship is quite intriguing.
Indeed, both analytical results in dust collapse
(see e.g. Jhingan \& Joshi 1997 and references therein)
and numerical results in scalar field
collapse (see e.g. Gundlach 1997 and references therein) 
indicate the existence of a critical behaviour
governing the formation of black holes or naked singularities.
In principle, this behaviour should be the ``remnant''
of some hypothesis of a - still unknown - cosmic censorship
theorem and, as such, should be related to the properties
of the collapsing matter 
like fulfilment of energy conditions and local stability.
If we want to understand the physics of such phenomena
in the case of ``ordinary'' matter (i.e.
not boson stars), 
we are enforced to approach analytically the gravitational collapse
of non pressureless matter, since the dust equation of state 
is ``trivial'' from the viewpoint of matter properties:
all energy conditions, causality conditions and stability conditions
reduce to positivity of energy.
However, this problem is extremely difficult
if approached in full generality (see e.g. Joshi 1996
and references therein).
\par
Recently (Magli 1997, to be referred towards as [I]), 
we discussed
a class of solutions of the Einstein field equations 
describing spherically symmetric, non--static elastic spheres
supported only by tangential stresses (elastic matter 
is of interest in strongly collapsed situations; for instance
neutron stars typically have solid regions - see e.g.
Haensel 1995).
Previous investigations on systems having vanishing radial stresses
trace back to Einstein (1939)
and Florides (1974) in the static case, while non--static models 
have been considered by
Datta (1970), Bondi (1971),
and Herrera \& Santos (1995) (for a general review on anisotropic systems
in General Relativity see Herrera \& Santos 1997).
\par
From the viewpoint of cosmic censorship,
understanding the nature of singularities 
for such solutions can be considered as a first step 
toward a full understanding of spherical gravitational collapse with 
general stresses.
Indeed the investigation of some particular models 
carried out recently by Singh and Witten (1997)
already shows behaviours which can be drastically different from the 
dust ones.
The purpose of this paper is to contribute to a 
program which should hopefully 
lead to a full understanding of the final fate of gravitational
collapse with tangential stresses.
\par
The paper is organized as follows.
The first problem we have to face with is 
the fact that, due to the presence of stresses,
the comoving time differs 
from the proper time of the shells of particles.
As a consequence, 
the first order ``energy equation'' which arises 
from mass conservation is coupled, for generic equations of state,
to the equation giving the acceleration of the world lines
of the particles.
This coupling has the effect that the equation for null geodesics
(whose behaviour near the central singularity governs the nature
of the collapse) cannot be written in explicit form.
Here, we overcome this difficulty (section II)
using mass--area coordinates. 
These coordinates
were originally introduced 
by Ori (1990) to obtain the general exact solution for charged dust.
Here we obtain the general solution
for gravitational collapse with non vanishing tangential stresses
in a very simple form,
in which only an integral
remains to be performed.
\par
Using the line element in the new coordinates, it is possible (section III)
to write the ``root equation'', namely the algebraic
equation governing the nature of the central singularity, 
in explicit form.
This equation depends only on the choice of the equation of state 
and of the initial distributions of density and velocity.
\par
It turns out
to be a quite difficult task to study this equation 
in full generality;
it is, however,
worth mentioning
that a full understanding of the nature of the singularities
for {\it dust} spacetimes
has been achieved
only very recently (Singh \& Joshi 1996).
To get some insights 
into the physical contents
of this equation,
we carry out a detailed examination of small deviations from the
dust equation of state.
This leads us to formulate a
conjecture on the final fate of the collapse 
and to give a plausibility
argument supporting it (section IV).
Finally, in section V, we analyze 
the case of self--similar spacetimes 
putting in evidence some qualitative 
effects of the tangential stress on the 
nature of the central singularity.
\par
The paper ends with some concluding remarks in section VI.

\pretitolo
{\bf II. Mass--Area coordinates and general solution }
\postitolo

Recently (see [I]) we discussed a 
class of solutions of the Einstein field equations 
describing spherically symmetric, non--static elastic spheres
supported only by tangential stresses.
Using comoving coordinates,
the line element reads
$$
ds^2=-e^{2\nu}dt^2 +\frac{(Y')^2 h^2}{1+f}dr^2 +Y^2
(d\theta^2 +\sin^2\theta d\phi^2)\ ,\autoeqno{lel}
$$
where $\nu$
and $Y$ are function of $r$ and $t$ 
satisfying
$$
\nu' = -\frac 1h \frac{\partial h}{\partial Y} Y' \ ,\autoeqno{coup1}
$$
$$
\dot Y^2 e^{-2\nu}
= -1 + \frac {2F}Y+\frac {1+f}{h^2}
\ .
\autoeqno{coup2}
$$
In the above formulae,
$F(r)$ and $f(r)$ are the ``conserved mass'' and 
the ``binding function'' familiar from the Tolman--Bondi solutions,
while $h=h(r,Y)$ is the internal elastic energy 
per unit volume
(it plays the role of equation of state of the material).
For physical reasonability, 
$F$, $1+f$ and $h$ must be chosen
as positive functions
(positivity of mass, 
of $g_{rr}$ and of elastic energy, respectively).
Moreover, the function $h$ is severely constrained 
by the local stability of matter, which requires 
this function to have a minimum at $Y=r$ (see [I] for details).
If $h$ is equal to one 
(more precisely, is a constant which may be rescaled to unity)
there is no dependence on the strain:
the material is a dust cloud and the line element
reduces to the Tolman--Bondi one.
This fact will be very important 
in what follows;
we shall call {\it dust limit} of any equation 
depending on the choice of $h$
the same equation written with $h=1$.
\par
The energy-momentum tensor of the material is diagonal
in the comoving frame and has only three
non--vanishing components,
namely the energy density 
$\epsilon=-T^0_0$ and the tangential stress $\Pi=T^\theta_\theta=T^\phi_\phi$.
These quantities 
are given by
$$
\epsilon =\frac {F'}{4\pi Y^2 Y'}
\ ,\autoeqno{en1}
$$
$$
\Pi = \hh\epsilon \ ,\autoeqno{enpi}
$$
where the ``generalized adiabatic index'' $\hh$ 
is defined as follows:
$$
\hh :=-\frac Y{2h} \diff hY\ .\autoeqno{defadi}
$$
\par
In [I] the physical properties 
of these solutions are thoroughly discussed.
We recall here 
that the metric is well behaved at the centre 
if the equation of state 
satisfies the ``minimal stability requirement''
($h$ has a minimum at $Y=r$) and the conditions
$Y(0,t) =0$, $f(0) =h^2(0,0)-1$ hold.
The behaviour of the energy density at $r=0$ is the same as that familiar
from the Tolman--Bondi models, namely $\epsilon$ is initially regular
if $F(r)$ is of the form $r^3\tilde F(r)$
with $\tilde F(0)<+\infty$.
Matching our solutions with the Schwarzschild vacuum 
is possible on any chosen boundary surface
$r=r_b$, provided that the value of $F$ at $r_b$
is identified with the Sch\-warz\-schild mass $M$.
However, the transformation between comoving and Schwarzschild 
coordinates is highly non--trivial.
The energy conditions lead to inequalities 
on the function $\hh$, and therefore to differential inequalities on the state
function $h$.
In particular {\it wec} holds
if $\hh \geq -1$.
Once this is satisfied,
{\it dec} requires $\hh \leq 1$, while
{\it sec}
is satisfied if $\hh \geq -1/2$.
Analyzing the behaviour of the function appearing
at the right hand side 
of equation \eqref{coup2} for a fixed shell 
of particles ($r=\const$), one can give a qualitative analysis
of the possible motions.
In particular, 
it is shown in [I] that
there exist physically valid models of
oscillating elastic spheres as well as of finite--bouncing spheres.
\par
The metrics \eqref{lel}
are not completely explicit due to the coupling between eqs. 
\eqref{coup1} and \eqref{coup2};
physically, this is simply a reflection of the fact
that the comoving time differs from the proper time 
since the particles are 
not in geodesic motion.
The two equations decouple only if
$h$ depends uniquely on $Y$ ($h=w(Y)$, say).
In this case we have
$$
\eqalign{
e^{2\nu}&=\frac 1{w^2}\ ,\cr
\dot Y^2 &=\frac 1{w^2}\left[
\frac {1+f}{w^2}-\left(
1-\frac {2F}Y\right)\right]\ ,
}
$$
and the function $\hh$ defined in \eqref{defadi}
reads
$$
\hh =-\frac Y2  \frac{d}{dY} \log w(Y)\ .
$$
This particular class of solutions contains that discussed by 
Singh \& Witten (1997), in which the tangential stress 
is proportional to the density.
In this case the stress--strain relation
has the ``barotropic form''
$\Pi=k\epsilon$ with constant ``adiabatic index'' $\hh =k$,
and the equation of state is 
$w(Y)=Y^{-2k}$.
\par
A consequence of the above described coupling problem 
is that, in general,
it is not possible to
write explicitly the null geodesic equation 
in comoving coordinates.
As will be recalled in the next section, this equation governs the nature
(naked or black hole) of the central singularity,
and it is therefore very difficult to investigate
on censorship using the comoving frame.
\par
A system which is mathematically
very similar to ours is charged,
spherically symmetric dust 
(Vickers 1973).
Indeed for such a system the mass is conserved
and it is possible to analyze the dynamics in comoving coordinates;
however the coupling between ``times'' does not allow explicit
integration.
To get rid of this problem
Ori (1990) introduced a system of coordinates 
which removes 
the coupling and obtained the general exact solution
for charged dust.
Ori's system is obtained 
by replacing the 
comoving time $t$ and the radial label $r$
with the ``area coordinate'' $R=Y(r,t)$ and the
``mass coordinate'' $m=F(r)$;
the mass coordinate, being conserved, is comoving
(if $u^\mu$ denotes the velocity of matter,
one has $u^\mu = u\id \mu{R}$ where $u:=u^R$).
The line element in these coordinates 
has the form
$$
ds^2=-Adm^2 -2BdRdm -C dR^2 +R^2 
(d\theta^2 + \sin^2\theta
d\varphi^2)\ ,
$$
where $A$, $B$ and $C$ are functions of $m$ and $R$.
\par
We are now going to show that Ori's technique can be applied 
also in the case of solutions with vanishing radial stresses.
This is essentially due 
to the fact that
the mass
is conserved 
also in this case,
and therefore gives an unambiguous ``comoving label''
for the shell of particles
(for simplicity, the general solution 
is presented here in the non--charged case;
however our results can be easily extended to the case of charged
materials, as briefly reported in appendix).
\par
Since $m$ is comoving, we have
$$
C=\frac 1{u^2}\ .
$$
The energy density can now be written as
$$
\epsilon = \frac {h}{4\pi uR^2E\H}\ .\autoeqno{ego}
$$
where $h$ is to be considered as a function of $m$ and $R$,
the quantity $H$ is defined by
$$
H := B^2-AC\ ,
$$
and $E$ is an arbitrary function of $m$
corresponding to $\sqrt{1+f}$
(we have introduced this notation
in order to facilitate the comparison with Ori's 1990 paper).
\par
There are four (compatible) 
Einstein equations for the three unknowns $A,B,C$.
We start considering equations $G^m_m=8\pi T^m_m$
$G_m^R=8\pi T_m^R$ and $G_R^m=8\pi T_R^m$:
$$
\frac 1{R^2}\left[
1-\frac AH -R\left(\frac AH\right)_{,R}
\right]
=0\ ,\autoeqno{mefe1}
$$
$$ 
\frac 1R \left(\frac AH\right)_{,m}=8\pi Bu^2\epsilon
\ ,\autoeqno{mefe2}
$$
$$
B\frac{H_{,R}}H -C_{,m}=0\ .
\autoeqno{mefe3}
$$
Equation \eqref{mefe1} can be integrated:
$$
A =H \left(1-\frac {2m}R\right) \ .\autoeqno{sol1}
$$
Using $B^2=H+A/u^2$ 
and equation \eqref{ego}, from \eqref{mefe2} 
we get
$$
u=\pm \sqrt{-1+\frac {2m}R +\frac{E^2}{h^2}}
\ ,\autoeqno{rw}
$$
and the metric function $B$ can be written as 
$$
B =-\frac {E\sqrt{H}}{hu}\ .
$$
Therefore,
we have solved for $u$ (and thus for $C$) in terms of the arbitrary functions 
and expressed $A$ and $B$ in terms of a single unknown $H$. 
To complete the solution, we plug the above results in \eqref{mefe3},
obtaining
$$
\left(\H \right)_{,R}
=\frac hE \left(\frac 1u\right)_{,m}\ .
$$
We have, therefore, reduced the problem
to the calculation of an indefinite integral:
$$
\sqrt{H (m,R)} =g(m) \pm \int
G(m,R)dR\ ,\autoeqno{hmr}
$$
where 
the $\pm$ sign is the same as that of $u$,
$g(m)$ is an arbitrary  function,
and
$$
G (m,R) :=
\frac {h}{RE} 
\left[1+\frac R2 \left(
\frac {E^2}{h^2}
\right)_{,m} \right]
\left(-1+\frac{2m}R+\frac {E^2}{h^2}\right)^{-\frac 32}
\ .\autoeqno{nonc}
$$
If $h=1$ the above formulae give
the Tolman--Bondi line element in mass--area coordinates
(Ori 1990).
\par
It is easy to check that the remaining 
field equation $G^\theta_\theta=8\pi T^\theta_\theta$
is identically satisfied once eqs.
\eqref{mefe1}, \eqref{mefe2} and \eqref{mefe3} are.
\par

\pretitolo
{\bf III. A laboratory for Cosmic Censorship}
\postitolo

In this section we use the general exact solution derived
in the previous section to build up a ``laboratory''
for studying cosmic censorship.
The key instrument which is needed in this laboratory
is already known from the work by Dwivedi \& Joshi (1994)
and may be called ``root equation''; as recalled below,
it is an algebraic equation arising from the behaviour of outgoing
null geodesics near the singularity.
In the present section
we construct this equation explicitly for the case at hand.
We shall also derive 
the conditions for shell--crossings singularities
in terms of an integral equation.

\pre
{\sl 3.1 Physical content of the arbitrary functions: initial data}
\post

In order to approach the problem of singularities,
we first need to identify 
the physical content of the arbitrary functions, 
so that regular initial data can be chosen.
\par
There are three arbitrary functions,
namely 
the equation of state $h=h(m,R)$
and the functions $E(m),g(m)$.
To understand the physical meaning of $E$ and $g $ observe that,
physically, such functions must be related to the 
``initial distributions'' of density and velocity
(here quotation marks are due to the fact
that we shall take care of the initial data always
referring to the ``original'' - comoving - coordinates).
Consider, therefore, regular initial data 
at some comoving time $t$ ($t=0$, say).
We use the scaling freedom in the choice of the $r$ coordinate to identify
the lagrangian and the eulerian label initially, so that $Y(r,0)=r$.
In mass--area coordinates,
to the equation $Y(r,0)=r$ corresponds some curve 
$R=\a(m)$, where $\a=F^{-1}$
and we are assuming the mass to be a monotonically increasing
function.
Introducing the initial distribution of velocity ($V(m)$, say)
from \eqref{rw} it follows
$$
V^2 (m)=
\frac {E^2(m)}{h^2(m,\a)}
-1+\frac {2m}{\a}\ .
$$
The above formula gives the relationship between $E$ 
and the initial velocity profile. 
\par
The relationship between $g $ and the initial data is, in general,
quite complicated.
To obtain it,
observe that the following formula may be easily proved:
$$
\diff Ym =\frac E{h} u\H\ .\autoeqno{ym}
$$
The above equation evaluated ``at $t=0$''
yields
$$
g  (m)=\pm\left[
\frac {\a_{,m}(m)h(m,\a)}{V(m)E(m)}
-
\int^{\a(m)}
G(m,R)dR\right]
\ ,\autoeqno{cfe}
$$
(the $\pm$ sign is the same as that of $u$).
\par

\pre
{\sl 3.2 Shell--focussing singularities}
\post

The energy density \eqref{en1} becomes singular whenever 
$Y(r,t)$ or $Y'(r,t)$ vanish during the dynamics.
Physically, such singularities correspond
to those occurring in dust models:
$Y=0$ corresponds to ``crushing to zero size''
(shell--focussing singularities) 
while $Y'=0$ 
corresponds 
to the 
shell crossing phenomenon:
the world lines
of the (shells of) particles intersect each other and the
``lagrangian labelling'' description breaks down.
\par
Contrary to what happens in the dust case,
where shell--focussing collapse in unavoidable,
within our solutions there exist globally regular
models of oscillating or bouncing back materials.
However,
equation \eqref{coup2}
can be used (see [I]) 
to show that 
for any physically valid choice of the equation of state
it is possible to choose initial data leading to continued 
gravitational collapse and therefore to shell--focussing
singularities
(it is worth mentioning that the remark
made by Singh \& Witten (1997) that regularity   
conditions ``explicitly disallow
the formation of a singularity at $r=0$''
is incorrect: these conditions imply only
regularity on the initial data surface and non--preferredness 
of the centre).
\par
To analyze the nature of shell--focussing
singularities,
we first observe that
the relation $Y(r,t)=0$ defines a ``singularity curve'' 
$t^{\rm s}(r)$; in general, different shells
become singular at different times,
and it is customary to call central singularity
that occurring at $r=0$.
This singularity plays a distinguished role because 
it is possible to show that
non--central singularities 
are always covered.
For, notice that
the shell labelled $r$ becomes trapped at a time $t^{\rm t}(r)$
such that $Y(r,t^{\rm t})=2F(r)$.
For each fixed shell, consider the function $\bar Y (t)=Y(r,t)$.
We have $\bar Y (t^{\rm t})=2F(r)>\bar Y (t^{\rm s})=0$,
but $d\bar Y (t)/dt$ is negative in a collapsing situation,
so that $\bar Y (t)$ is decreasing and it must be $t^{\rm s}(r)
>t^{\rm t}(r)$.
It follows that the shell becomes 
trapped before becoming singular,
so that the singularity is covered
(under certain conditions,
it is possible to proof this result also
in presence of non--vanishing radial stresses,
see Cooperstock {\it et al} (1997) for details).
\par
The above argument does not work for the central singularity,
at which $Y(0,t^{\rm t})=2F(0)=0$.
To study this singularity, we translate in the mass--area formalism
and make use of the method developed 
by Dwivedi \& Joshi (1994)
and successfully applied to the dust case 
in a series of recent papers
(see e.g. Singh \& Joshi 1996,
Jhingan {\it et al.} 1996).
\par
Consider the equation for radial, 
outgoing null geodesics
in mass--area coordinates:
$$
\frac {dR}{dm}=-\frac {B+\sqrt{B^2-AC}}C =-\sqrt{H}u\left(|u|-\frac Eh \right)
\ .\autoeqno{li}
$$
This is
an ordinary differential equation
with a singular point at the central singularity $R=0$, $m=0$.
This singularity is (at least locally) naked 
if there are geodesics starting at it
with a definite value of the tangent.
If no such geodesics exists, the singularity is not naked
and (strong) cosmic censorship holds.
To investigate the behaviour near the singular point,
define
$$
x:=\frac R{2m^\alpha}\ ,
$$
where $\alpha >1/3$.
If the singularity is naked,
there exist some $\alpha$ such that at least one finite positive
value $x_0$ exists which solves the algebraic equation
$$
x_0 :=\lu x =\lu \frac R{2m^\alpha}\ .
$$
Applying L'Hospital rule we have
$$
x_0 
=
\lu \frac {m^{1-\al}}{2\al} \frac {dR}{dm}
=
\lu
-\frac {m^{1-\al}}{2\al}
\sqrt{H}u
\left(|u|-\frac Eh \right)
\ .
$$
Using eqs. \eqref{rw} and \eqref{hmr}, 
the above equation can be written in explicit form as 
$$
\eqalign{
x_0 
=
&
\frac 1{2\al}
\lu  
m^{\frac 32(1-\al)} 
\left[
g  (m) -
\int^{2m^\al x}
G(m,R)
dR
\right]
\left(
\sqrt{
\left(-1+\frac {E^2}{h^2}\right)
m^{\al -1}+\frac 1x}\right)
\times
\cr
&\times
\left(
\sqrt{\frac {m^{1-\al}}x -1+\frac {E^2}{h^2}
}-\frac Eh\right)
\ .
}\autoeqno{lpsi}
$$
This equation
depends only on the initial data 
$g$ and $E$, as in the dust case,
and on the choice of the material we are dealing with,
i.e. the equation of state $h$.
This means that the dynamics has been 
completely ``gauged away'';
such a simplification cannot be achieved
in comoving coordinates since 
in such coordinates the general exact solution is not
available in explicit form.
In the next two sections, we shall illustrate a simple way
to extract physically interesting
information from this equation without solving it explicitly.
\par

\pre
{\sl 3.3 Remarks on shell--crossing singularities}
\post

As recalled above, shell crossing singularities correspond 
to zeroes of $Y'$, so that
in the mass--area description 
a shell crossing occurs
when $Y_{,m}$ vanishes.
Generally speaking, we do not expect a zero of $Y_{,m}$ to occur
at a turning point ($u=0$) so that equation \eqref{ym} 
implies that shell crossing singularities correspond to zeroes
of $H$.
Let $R^{\rm sc}(m)$ be the curve on which such singularities 
eventually occur.
Using eqs. \eqref{hmr} and \eqref{cfe}
we obtain that $R^{\rm sc}(m)$ must satisfy
to
$$
\frac {\a_{,m}(m)h(m,\a)}{V(m)E(m)}
=
-
\int_{\a(m)}^{R^{\rm sc}(m)}
G(m,R)dR\ .\autoeqno{rsc}
$$
This equation {\it may} have physically meaningful
solutions.
For instance, consider the Tol\-man---Bondi case.
The solutions of equation \eqref{rsc} 
are physically meaningful only if the vanishing of $Y_{,m}$
happens {\it before} (in comoving time terms) the singularity
at $Y=0$ is reached. 
It is possible to characterize fully in terms of differential inequalities
the set of initial data such that no shell crossing occur 
in physically allowed ``times'' (Hellaby \& Lake 1985,
Newman 1986, Jhingan \& Joshi 1998).
On the contrary, Ori (1990) used the charged dust counterpart
of this equation (see appendix)
to show that the characteristic 
``bounce in a new universe'' process (De La Cruz \&
Israel 1967) which is typical in such solutions 
always occur {\it after} a shell--crossing, thereby casting serious
doubts on its physical realizability (see also Ori 1991).
\par
In the general case of non--vanishing tangential stresses,
the analysis is also possible in full generality and will be 
presented elsewhere.

\pretitolo
{\bf IV. The nature of the central singularity: a conjecture}
\postitolo

As we have seen, 
the nature of the central singularity
depends on the existence of solutions of equation
\eqref{lpsi}.
A complete study of this equation requires a detailed
investigation on the behaviour of the 
equation of state in the limit of approach to the singularity
in physically valid situations, and goes
far beyond the scope of the present paper.
However, 
some insights into this problem can be 
obtained by a careful analysis and comparison with
the (already well known) results holding for
dust.
For our considerations it will be sufficient to 
consider the case 
of marginally bound collapse;
we are, therefore, going
to give a simple derivation in mass--area
coordinates of the results on the nature of the central singularity
in this case (we completely refer the reader
to the original paper by Singh \& Joshi (1996) for details).
\par
The marginally bound
dust case corresponds to $E=h=1$.
From formula \eqref{cfe} we get 
$$
g  (m) = \pm\sqrt{\frac {\a}{2m}}m^{\tre}\frac{d}{dm}(m^{-\tre}\a).
\autoeqno{psie}
$$
The behaviour of this function as $m$ tends to zero
can be obtained as follows.
Consider regular initial data in comoving coordinates.
Then the function $F$
will be of the form
$$
F(r)= F_0 r^3 + F_q r^{q+3} +...
$$
where $q$ is the order of the 
first non vanishing derivative of the initial density profile
at the centre, and dots stand for higher order terms.
Therefore, we have
$$
\a (m)=F^{-1} (m)= \left(\frac m{F_0}\right)^{\tre} 
-\frac {F_q}{3F_0}\left(\frac m{F_0}\right)^{\frac{1+q}3}+...
$$
Considering now equation \eqref{psie} and recalling that 
we are considering collapse (so that the negative sign must be chosen)
we obtain
$$
g  \approx P_q m^{\frac q3 -1}\ ,
$$
where 
$$
P_q:=\frac{qF_q}{
9\sqrt{2}F_0^{\frac q3+\frac 32}}
\ .
$$
Thus $g $ exhibits a ``critical'' behaviour:
it diverges (respectively, goes to a finite non--zero limit,
vanishes) if $q<3$, $q=3$, $q>3$.
Surprisingly enough,
it is this behaviour that governs 
the nature of the singularity.
In fact,
equation \eqref{lpsi} yields
$$
x_0 
=
\frac 1{2\al}
\lu 
\left[
P_qm^{
\frac q3 +\due (1-3\alpha) 
}
-\frac 23 
x^{\frac 32}
\right]
\frac 1{\sqrt{x}}
\left(
\frac {m^{ \frac{1-\al}{2} } }{\sqrt{x}}
-1\right)
\ .
$$
The first term in square brackets goes to a finite, non zero limit iff
$$
\al =\tre \left(1+\frac {2q}3\right)\ ,
$$
so we get
$$
x_0 
=
\frac 1{\frac 23 \left(1+\frac{2q}3\right) }
\left(
P_q-\frac 23 
x_0^{\frac 32}
\right)
\frac 1{\sqrt{x_0}}
\lu
\left(
\frac{m^{ \tre \left(
1-\frac q3 \right) }}{\sqrt{x}}
-1\right)
\ .
\autoeqno{dive}
$$
If $q$ is ``super--critical'' ($q>3$)
the limit diverges: there are 
no null geodesics escaping 
and therefore the singularity is not naked.
If $q$ is ``sub--critical'' ($q=1,2$) 
the limit goes to minus one
and \eqref{dive} gives a real
positive solution for 
$x_0$ (provided that the initial density is decreasing outwards):
the singularity is naked.
At the critical value $q=3$ \eqref{dive} becomes 
a quartic equation.
This equation has no real positive roots 
(and therefore the singularity is covered)
if the quantity $\zeta=F_3/(2\sqrt{2}F_0^{5/2})$
is greater than a certain numerical value,
otherwise the singularity is naked.
\par
The qualitative features of the general (i.e. non--marginally bound) case
are similar to those recalled above, namely nakedness
depends on the ``critical'' behaviour of some
parameter $\tilde q$ (which reduces to the parameter $q$ 
in the marginally bound case);
if the singularity is censored the limit diverges for any
$\tilde q$ greater than the critical value
(Singh \& Joshi 1996).
\par
Consider now a generic solution with tangential stresses.
To identify it uniquely, 
we need to chose the equation of state and 
the initial distribution of density and
velocity.
This means that the space of the free functions
can be visualized as follows: to any fixed choice of the initial data
$g(m)$ and $E(m)$ corresponds a family of solutions 
$\s_h$.
Each member of this family corresponds to a different material
(a different choice of $h(m,R)$ within the physically 
allowed range) and each family contains one and only one 
Tolman--Bondi solution $\s_1$ (``dust limit'') corresponding to 
$h=1$.
Chosen a family $\s_h$, we can immediately infer 
from the Singh \& Joshi work if the dust limit $\s_1$ 
corresponds to a naked singularity or to a blackhole.
\par
We conjecture that, if the central singularity
of $\s_1$ is not naked and it is not critical
(i.e., the limit in \eqref{dive}
is divergent), the central singularity 
of $\s_h$ is also not naked for any physically valid choice of $h$.
Roughly speaking, this means that one cannot use a
physically valid tangential stress to undress a covered 
dust singularity.
\par
The above conjecture is based on arguments of physical
plausibility as follows.
Consider a small deviation from the dust equation of state.
This can be represented as 
$$
h=1+\mu (R,m) \ , \autoeqno{hmu}
$$ 
where the function
$\mu$ is positive and
vanishes at $R=\a (m)$
(a reasonable choice for $\mu$ could be the ``quasi--hookean''
equation of state
$\mu =\mu_0 (m) (R-\a)^2$
with positive $\mu_0$).
Then each term in equation 
\eqref{lpsi} can be expanded to first order in $\mu$;
if this function is physically valid
(i.e. is chosen as described above) every such terms
will be quadratic in $R-\a$.
In particular, the last factor in round brackets 
will have this behaviour.
Now, it is easy to check
that quadratic terms in $R-\a$
(or higher order terms)
cannot regularize a diverging behaviour of the zero--order
term in this factor. Since divergency occurs
in the non critical covered case, 
the conjecture is proved at least for small deviations 
from the dust equation of state.
\par

\pretitolo
{\bf V. Self--similar spacetimes}
\postitolo

The above described proposal on the nature 
of the central singularity, although covering a large sector of the space of 
initial data, leaves 
completely open the problem of interpretation of critical behaviour.
For instance:
what happens to marginally bound 
solutions having a naked dust limit with $q=3$?
Do such solutions remain naked with the addition of {\it any}
(physically valid) tangential stress?
Do they become {\it always} covered? 
It seems likely that none of the above would hold, but rather 
that the threshold of black hole formation 
for fixed initial data should depend on the equation of state
(i.e. on the choice of the function $h$),
hopefully in a physically reasonable and understandable way. 
We don't have the answer to this question yet.
However, we are going to present here some 
(again, qualitative) evidence that really a behaviour 
like this should occur.
\par
Difficulties in studying critical cases arise 
because we must investigate finite values of the limit
\eqref{lpsi} and, therefore, existence of positive solutions 
of the root equation. 
To get some insight into this 
we consider, among the solutions presented above,
a particulary simple case which, however,
is not deserved of physics,
namely the case of self--similar spacetimes
(see e.g. Carr 1997).
\par
In mass area coordinates we can use as self similar variable
the quantity $x=R/2m$.
It is easy to check that
the spacetimes are self--similar
if the following conditions hold
(see Magli 1993 for a discussion of self--similarity 
in the case of anisotropic matter):
\capo
1) $h$ is a function of $x$ only
\capo
2) $g(m)$ and $E(m)$ are constant.
\capo
In what follows, 
it will again be sufficient
to consider the marginally bound case $E=1$.
\capo
The above conditions imply that $\a (m)$ is a linear function,
and indeed it is well known that for self--similar
spacetimes the mass function (the inverse of $\a$) is linear.
We therefore set $F(r)=\lambda r/2$ ($\lambda =\const$) so that
$$
\a (m) =\frac {2m}\lambda \ ,
$$
this implies that the value $\overline x$ of $x$ at initial data is $1/\lambda$.
Using equation \eqref{cfe} with $h(\overline x )=1$, 
the root equation \eqref{lpsi} can be written as
$$
x 
=
-
\frac 1{2}
\left[
\frac 1{6\ala}
+
\int_{\overline x}^x
{\cal G}(z)
dz
\right]
\sqrt{
-1+\frac {1}{h^2}
+
\frac 1x}
\left(
\sqrt{
-1+\frac {1}{h^2}
+
\frac 1x}
-
\frac 1h
\right)
\ ,
\autoeqno{lpsiss}
$$
where
$\ala :=\lambda^{\frac 32}/{12}$ and
$$
{\cal G}(x)
:=
\frac {h}x
\left(1+2\frac{x^2}{h^3}\di hx\right)
\left(-1 +\frac 1{h^2} +\frac 1x\right)^{-\frac 32}
\ .\autoeqno{hatg}
$$
\par
To extract from equation \eqref{lpsiss} 
some qualitative information 
we again start 
from the dust limit.
What happens in this limit is already well known
(Joshi \& Singh 1995),
and we completely refer the reader to this paper for details.
\par
Setting $h=1$ equation \eqref{lpsiss} simplifies to
$$
x 
=
-
\frac 1{2}
\left(
\frac 1{9\ala}
+
\frac 23 x^{3/2}
\right)
\sqrt{
\frac 1x}
\left(
\sqrt{
\frac 1x}
-1
\right)
\ .
\autoeqno{lpsissd}
$$
To facilitate the comparison with the Joshi \& Singh paper 
we change variable to
$$
y:=\frac 23 \left(1-12\ala x^{\frac 32}\right)
$$
then \eqref{lpsissd} becomes a quartic equation in the variable $y$:
$$
y^3\left(\frac 23-y\right)=\ala (2-y)^3
\ ,
\autoeqno{come}
$$
where $0<y<2/3$.
Rewriting this equation in canonical form
as $ay^4 +4b y^3 +6c y^2 +4d y +e =0$, it can be shown 
that real, positive solutions exist
only if the quantity
$$
\Delta:=
(ae-4bd+3c^2)^3-27
(ace+2bcd-ad^2-eb^2-c^3)^2
\ ,
$$
is negative (in this case there are two such solutions).
This happens if $\ala<\ala_1\circa 6.41 \times 10^{-3}$ 
or $\ala>\ala_2\circa 17.32$.
Thus the collapse leads to 
black hole formation
if $\ala_1 <\ala <\ala_2$, to naked singularities otherwise.
The range $\ala>\ala_2$ is, however, unphysical
since it would correspond to imaginary values of $x$.
Therefore self--similar dust spacetimes exhibit a ``phase transition''
between naked singularities
and black holes.
The transition depends on the value of $\ala$ which is the remaining
free parameter.
This quantity is related
to the central density of the material;
using a fiducial model Joshi \& Singh have shown that $\ala$
typically do belong
to the range of black hole formation
for densities near the nuclear one.
\par
We now want to investigate, at least qualitatively,
the changes introduced in the above picture 
by the presence of tangential stresses.
For, consider once again a small deviation
from the dust equation of state of the form \eqref{hmu}
(obviously, for self--similar spacetimes
$\mu$ has to be be considered a function of $x$
only).
In what follows, 
we shall systematically discard terms of order higher than one
in $\mu$.
Expanding the root equation \eqref{lpsiss}
we obtain
$$
y^3\left( y-\frac 23\right)=\ala (y-2+\tilde K)^3\ .\autoeqno{come1}
$$
In the above formula, $\tilde K$ is the value at order one
of the following function
$$
\eqalign{
K(y):=
&
\frac 9{2y}
\left(\frac 23 -y\right)
\left[
\frac{(2-y)
\left(\frac 23 -y\right)
}{4y^2}
\mu
+
\ala
\int^{
\frac 1\lambda
}_{
\frac 1\lambda
\left(1-\frac 32 y\right)}
\sqrt{z}\left(
(1+3z)\mu+2z^2\di \mu{z}\right)dz
\right]
\ .
}\autoeqno{kti}
$$
Notice that this function is strictly positive.
Linearity in $\mu$ also implies 
$\tilde K=K(y_0)$ where $y_0$ is the solution
of the dust quartic \eqref{come}
in the neighbourhood of which we want to study 
the deviation from the dust case.
\par
Expanding also
equation \eqref{come1}  to first order,
we finally obtain a quartic with ``displaced''
parameters $a,b,c,d,e$.
A quite long, but straightforward, calculation
gives that the range of black hole formation $\ala_1 <\ala (<\ala_2)$
is altered 
by the perturbation as
$\tilde\ala_1 <\ala (<\tilde\ala_2)$,
where
$$
\eqalign{
\tilde \ala_1 &=
\ala_1 -\tilde K \delta_1 \ ,\cr
\tilde \ala_2 &=
\ala_2 +\tilde K \delta_2
}\autoeqno{eqw}
$$
and $\delta_1 \circa 17\times 10^{-3}$, $\delta_2 \circa 12.8$
(as usual in any perturbative approach, the above results
also give the condition of applicability of 
the approximation: $\tilde K$ must be not greater
than about $1/3$).
\par
Formulae \eqref{eqw}
show, at least at a qualitative level,
that black hole formation is facilitated 
by the presence of tangential stresses.
Indeed, since $\tilde K$ is positive,
the upper bound $\ala_2$ 
becomes higher and certainly remains unphysical,
while the lower bound $\gamma_1$ 
tends to decrease.
It might happen that the addition of tangential stress 
{\it dresses} the singularity, which would be equivalent
to $\tilde\ala_1 <0$, but of course to draw a conclusion of this kind
it will be necessary to investigate the root equation without 
approximations.

\pretitolo
{\bf VI. Concluding remarks}
\postitolo

The results of the present paper can be summarized as follows.
First of all,
we have shown that Ori's mass--area formalism
can be used to 
bring the general spherically symmetric 
solution of the Einstein field equations
with non--vanishing tangential stresses
in a very explicit form, in which only the calculation of an integral
remains to be performed
(this can be done independently whether
electromagnetic coupling is present or not;
the generalization to charged materials
is indeed straightforward and is briefly reported in the appendix).
\par
The introduction of mass--area coordinates 
proves to be
a very powerful tool 
as far as the analysis on existence and nature of singularities
in such solutions is concerned.
In fact
it allows
to obtain the root equation governing the nature of the central singularity
in explicit form.
\par
We presented first results 
coming from the investigation of the dust limit 
of this equation.
Such results 
give some insight about what should be 
the nature of the final fate 
of collapse with tangential stresses.
In particular, we proposed 
a conjecture which, roughly speaking,
asserts that ``tangential stress cannot undress covered dust''.
In the last section, we discussed
self--similar spacetimes and
showed, at least qualitatively,
that the effect of the stress can be an enlargement of
the blackhole initial data space.
\par
Both the above recalled results 
depend on the structure of the state equation,
and therefore show
once again and in a clear way that
a connection should exist between a (still lacking)
mathematically rigorous formulation of cosmic censorship 
and the conditions of physical acceptability of the equations of state.
Such conditions obviously include 
the energy conditions but 
also the existence of an absolute
minimum of the internal energy, which is intimately related 
to stability issues;
a relevant improvement in our understanding 
of this topic could come from the knowledge of the 
explicit structure of the blackhole threshold
in terms of the derivatives of the state equation
evaluated near the singularity.
Work in this direction is now in progress. 

\pretitolo
{\bf Acknowledgments}
\postitolo

The final version of this paper benefited very much of many discussions
had by the author during a visit at the Theoretical Astrophysics Group,
Tata Institute of Fundamental Research, Bombay, India.
The author expresses his thanks to all members of this group,
and in particular to
S. Jhingan, P.S. Joshi and T.P. Singh.
The author also acknowledges Elisa Brinis Udeschini 
and T. Harada for useful discussions.


\pretitolo
\centerline{\bf References}
\postitolo

\vskip .1truecm 
\capo
Bondi, H., (1971)
``On Datta's spherically symmetric systems in General Relativity.''
Gen. Rel. Grav. {\bf 2}, p.321.
\vskip .1truecm 
\capo
Carr, B.J. (1997)
``Spherically symmetric similarity
solutions and their applications in Cosmology and Astrophysics.''
To appear in ``Proceedings of the 7$^{\rm th}$ canadian conference
on General Relativity and Relativistic Astrophysics''
\vskip .1truecm 
\capo
Cooperstock, F.J., Jhingan, S., Joshi, P.S., Singh, T.P., (1997)
``Cosmic censorship and the role of pressure in gravitational collapse.''
Class. Quantum Gravity, {\bf 14} 2195.
\vskip .1truecm 
\capo
Datta, B.K., (1970) 
``Non--static spherically symmetric clusters of particles
in General Relativity:I.''
Gen. Rel. Grav. {\bf 1}, p.19.
\vskip .1truecm 
\capo
De La Cruz, V., Israel, W., (1967)
``Gravitational bounce.''
Nuovo Cimento {\bf 51A}, p.744.
\vskip .1truecm 
\capo
Dwivedi, I.H., Joshi, P.S., (1994)
``On the occurrence of naked singularities in spherically symmetric
gravitational collapse.''  
Comm. Math. Phys. {\bf 166}, p.117.
\vskip .1truecm 
\capo
Einstein, A., (1939) 
``On a stationary system with spherical symmetry
consisting of many gravitating masses.''
Ann. Math. {\bf 40}, 4, p.922.
\vskip .1truecm 
\capo
Florides, P.S., (1974) 
``A new interior Schwarzschild solution.''
Proc. R. Soc. L. {\bf A 337}, p.529.
\vskip .1truecm 
\capo
Gundlach, C., (1997)
``Critical phenomena in gravitational collapse.''
gr-qc/9712084,
To appear in Adv. Theor. Math. Phys.
\capo
Haensel, P., (1995)  
``Solid interiors of neutron stars and gravitational radiation.''
In {\it Astrophysical sources of gravitational radiation},
J.A. Marck and J.P. Lasota ed. (Les Houches 1995).
\vskip .1truecm 
\capo
Hellaby, C., Lake, K., (1985)
``Shell crossing and the Tolman model.''  
Ap. J. {\bf 290}, p.381.
\vskip .1truecm 
\capo
Herrera, L., Santos, N., (1997)
``Energy content of a slowly collapsing gravitating sphere.''
Gen. Rel. Grav. {\bf 27} p.1071.
\vskip .1truecm 
\capo
Herrera, L., Santos, N., (1997)
``Local anisotropy in self--gravitating systems.''
Phys. Rep. {\bf 286}, 2.
\vskip .1truecm 
\capo
Jhingan, S., Joshi, P.S. (1998)
``The final state of collapse in Einstein theory of gravitation.''
In ``Proceedings of the Haifa workshop on 
the  Internal Structure of Black Holes and 
Spacetime Singularities" Lior Burko and Amos Ori ed., 
Annals of the Israel Physical Society {\bf 13} (I.O.P., Bristol).
\vskip .1truecm 
\capo
Jhingan, S., Joshi, P.S., Singh, T.P. (1996)
``The final fate of spherical inhomogeneous dust collapse II.
Initial data and causal structure of the singularity.''  
Class. Quantum Gravity {\bf 13}, p.3057.
\vskip .1truecm 
\capo
Joshi, P.S. (1996)
{\it Global aspects in gravitation and cosmology.} 
Oxford Science Publ. (Clarendon, Oxford).
\vskip .1truecm 
\capo
Joshi, P.S., Singh, T.P. (1995) 
``Role of initial data in the gravitational collapse
of inhomogeneous dust.''
Phys. Rev. {\bf D51}, p.6778.
\vskip .1truecm 
\capo
Magli, G., (1993)
``The dynamical structure of the Einstein equations for a non rotating star.''
Gen. Rel. Grav. {\bf 25}, p. 441.
\vskip .1truecm 
\capo
Magli, G. (1997)
``Gravitational collapse with non--vanishing tangential stresses:
a generalization of the Tolman--Bondi model.''  
Class. Quantum Gravity {\bf 14}, p. 1937 (Paper [I]).
\vskip .1truecm 
\capo
Newman, R.P.A.C., (1986)
``Strengths of naked singularities in Tolman--Bondi spacetimes.''  
Class. Quantum Gravity {\bf 3}, p.527
\vskip .1truecm 
\capo
Ori, A. (1990)
``The general exact solution for spherical charged dust.''  
Class. Quantum Gravity {\bf 7}, p.985.
\vskip .1truecm 
\capo
Ori, A. (1991)
``Inevitability of shell crossing in the gravitational collapse
of weakly charged dust spheres.''  
Phys. Rev. {\bf D44}, p.2278.
\vskip .1truecm 
\capo
Singh, T.P., Joshi, P.S., (1996)
``The final fate of spherical inhomogeneous dust collapse.''  
Class. Quantum Gravity {\bf 13}, p.559.
\vskip .1truecm 
\capo
Singh, T.P., Witten, L. (1997)
``Spherical gravitational collapse
with tangential pressure.''  
Class. Quantum Gravity {\bf 14}, p.3489.
\vskip .1truecm 
\capo
Vickers, P.A. (1973)
``Charged dust spheres in General Relativity.''
Ann. Inst. Henri Poincar\'e {\bf 18}, p. 137.
\vskip .1truecm 
\capo
\vfill\eject

\pretitolo
{\bf Appendix: the charged case}
\postitolo

Consider a material carrying a non--vanishing 
charge density $\sigma$.
We keep the description of the mechanical and gravitational degrees of
freedom as in the body of the paper,
and simply
introduce the
Maxwell tensor which, due to spherical symmetry, has only one independent 
component $F^{mR}:=\Epsilon (m,R)$.
Maxwell's equations yield
$$
\eqalign{
\Epsilon &=\frac {Q}{R^2 \H}
\ ,
\cr
\sigma   &=\frac {Q_{,m}}{4\pi R^2 \H u} \ .
}
$$
where $Q=Q(m)$ is arbitrary.
The field equation \eqref{mefe1} now has a source term
and reads
$$
\frac 1{R^2}\left[1-\frac AH -R\left(\frac AH\right)_{,R}
\right]
=\frac {Q^2}{R^4}\ ,
$$
while
\eqref{mefe2} and \eqref{mefe3} remain unchanged.
Since $Q=Q(m)$, the equation above can be integrated at once and gives
$$
A =H
\left(1-\frac {2m}R +\frac {Q^2}{R^2}\right)
\ .
$$
Using $B^2=H+A/u^2$ 
and equation \eqref{ego}, from \eqref{mefe2} 
we get
$$
u=\pm 
\sqrt{
-1
+\frac {2m}R -\frac {Q^2}{R^2}+
\frac 1{h^2} \left(E-\X \frac QR\right)^2 
}
\ ,
$$
where the ``specific charge''
$\X :=EQ_{,m}$. 
The metric function $B$ can be written as
$$
B =-\frac 1{hu}
\left(E-\X \frac QR\right) \sqrt{H}\ ,\
$$
so that \eqref{mefe3} gives
$$
\left(\H \right)_{,R}
=\frac h{E-\X \frac QR}  \left(\frac 1u\right)_{,m}\ .
$$
It follows
$$
\sqrt{H (m,R)} =g (m) \pm \int
\widetilde G (m,R)dR\ ,\
$$
where $g(m)$ is arbitrary and 
$$
\eqalign{
\widetilde G 
&:=
\frac{h}{R\left(E-\X \frac QR\right)
}
\left\{
1-\frac{QQ_{,m}}{R}+\frac R2 
\left[\frac 1{h^2}\left(E-\X \frac QR\right)^2\right]_{,m}
\right\}
\times \cr
&
\times
\left[
-1+\frac {2m}R -\frac {Q^2}{R^2}
+
\frac 1{h^2} \left(E-\X \frac QR\right)^2 
\right]^{-\frac 32}\ .
}
$$
For $h=1$ (dust case) the function $\widetilde G$ 
is a rational fraction and its integral may be carried out
explicitly  (Ori 1990), while for $Q=0$ (non--charged case)
the above function coincides with that defined in \eqref{nonc}.

\end